\input psfig.sty
\documentstyle[epsf]{aipprocsp}
\def\vev#1{\langle #1 \rangle}
\def\lam{\lambda}

\def\lampr{\lam^\prime}
\def\lampp{\lam^{\prime\prime}}

\def\mplanck{M_{\rm Planck}}

\def\sq{\wt q}

\def\msq{m_{\sq}}
\def\slep{\wt \ell}

\def\mslep{m_{\slep}}
\def\slepl{\wt \ell_L}
\def\mslepl{m_{\slepl}}
\def\slepr{\wt \ell_R}
\def\mslepr{m_{\slepr}}

\def\sel{\wt e}

\def\sell{\wt e_L}

\def\selr{\wt e_R}

\def\cpmtwo{\wt \chi^{\pm}_2}

\def\delgs{\delta_{GS}}

\def\Twoloop{Two-loop/RGE-improved}

\def\slash#1{#1\hskip-8pt/\hskip5pt}

\def\etmiss{\slash E_T}
\def\rslash{R\hskip-5pt / \hskip5pt}
\def\susyslash{\susy\hskip-20pt/\hskip15pt}

\def\mhalf{m_{1/2}}
\def\gl{\wt g}
\def\mgl{m_{\gl}}

\def\stop{\wt t}
\def\stopone{\wt t_1}

\def\mstop{m_{\stop}}

\def\mstopone{m_{\stopone}}

\def\sbot{\wt b}

\def\sq{\wt q}

\def\msq{m_{\sq}}
\def\slep{\wt \ell}

\def\mslep{m_{\slep}}
\def\slepl{\wt \ell_L}
\def\mslepl{m_{\slepl}}
\def\slepr{\wt \ell_R}
\def\mslepr{m_{\slepr}}

\def\msusy{m_{\rm SUSY}}
\def\msusyslash{m_{\susy\hskip-15pt/\hskip15pt}}
\def\susy{{\rm SUSY}}

\def\etc{{\it etc.}}

\def\sign{{\rm sign}}

\def\etc{{\em etc.}}

\def\eg{{\it e.g.}}
\def\etal{{\it et al.}}
\def\mhalf{m_{1/2}}
\def\gl{\wt g}
\def\mgl{m_{\gl}}
\def\stop{\wt t}
\def\stopone{\wt t_1}
\def\mstop{m_{\stop}}

\def\mstopone{m_{\stopone}}
\def\sq{\wt q}

\def\msq{m_{\sq}}
\def\slep{\wt \ell}

\def\mslep{m_{\slep}}
\def\slepl{\wt \ell_L}
\def\mslepl{m_{\slepl}}
\def\slepr{\wt \ell_R}
\def\mslepr{m_{\slepr}}
\def\sbot{\wt b}

\def\hsm{h_{\rm SM}}

\def\hl{h^0}
\def\hh{H^0}
\def\ha{A^0}
\def\hp{H^+}
\def\hm{H^-}
\def\hpm{H^{\pm}}
\def\mhl{m_{\hl}}
\def\mhh{m_{\hh}}
\def\mha{m_{\ha}}

\def\mhpm{m_{\hpm}}
\def\tanb{\tan\beta}

\def\mz{m_Z}
\def\mw{m_W}
\def\mgut{M_U}
\def\mstring{M_S}

\def\wpm{W^{\pm}}

\def\chitil{\wt\chi}
\def\cnone{\wt\chi^0_1}
\def\cntwo{\wt\chi^0_2}

\def\snu{\wt\nu}

\def\snue{\wt\nu_e}
\def\msnue{m_{\snue}}
\def\snuel{\wt\nu_{e\,L}}

\def\mcnone{m_{\cnone}}
\def\mcntwo{m_{\cntwo}}

\def\h{h}

\def\wt{\widetilde}
\def\cpone{\wt \chi^+_1}
\def\cmone{\wt \chi^-_1}
\def\cpmone{\wt \chi^{\pm}_1}

\def\mcpmone{m_{\cpmone}}

\def\cpmtwo{\wt \chi^{\pm}_2}

\def\MPL #1 #2 #3 {Mod.~Phys.~Lett.~{\bf#1},\  #2 (#3)}
\def\NPB #1 #2 #3 {Nucl.~Phys.~{\bf#1},\  #2 (#3)}
\def\PLB #1 #2 #3 {Phys.~Lett.~{\bf#1},\  #2 (#3)}
\def\PR #1 #2 #3 {Phys.~Rep.~{\bf#1},\ #2 (#3)}
\def\PRD #1 #2 #3 {Phys.~Rev.~{\bf#1},\  #2 (#3)}
\def\PRL #1 #2 #3 {Phys.~Rev.~Lett.~{\bf#1},\  #2 (#3)}
\def\RMP #1 #2 #3 {Rev.~Mod.~Phys.~{\bf#1},\  #2 (#3)}
\def\ZP #1 #2 #3 {Z.~Phys.~{\bf#1},\  #2 (#3)}
\def\IJMP #1 #2 #3 {Int.~J.~Mod.~Phys.~{\bf#1},\  #2 (#3)}

\def\wpm{W^{\pm}}
\def\hpm{H^{\pm}}

\def\call{{\cal L}}
\def\wtil{\widetilde}
\def\what{\widehat}

\def\lam{\lambda}
\def\br{BF}

\def\gam{\gamma}

\def\etal{{\it et al.}}
\def\etc{{\it etc.}}

\def\anti{\overline}
\def\epem{e^+e^-}
\def\zstar{Z^\star}

\def\mupmum{\mu^+\mu^-}

\def\rts{\sqrt s}
\def\ie{{\it i.e.}}
\def\eg{{\it e.g.}}

\def\anti{\overline}

\def\mw{m_W}
\def\mz{m_Z}
\def\h{h}

\def\hsm{h_{SM}}

\def\tanb{\tan\beta}
\def\hl{h^0}
\def\mhl{m_{\hl}}

\def\ha{A^0}
\def\mha{m_{\ha}}

\def\hh{H^0}
\def\mhh{m_{\hh}}

\def\fbi{~{\rm fb}^{-1}}

\def\pbi{~{\rm pb}^{-1}}
\def\pb{~{\rm pb}}
\def\kev{~{\rm keV}}

\def\gev{~{\rm GeV}}
\def\tev{~{\rm TeV}}
\def\stop{\widetilde t}
\def\mstop{m_{\stop}}

\def\overlay#1#2{\ifmmode \setbox 0=\hbox {$#1$}\setbox 1=\hbox to\wd 0{\hss
$#2$\hss }\else \setbox 0=\hbox {#1}\setbox 1=\hbox to\wd 0{\hss #2\hss }\fi
#1\hskip -\wd 0\box 1}
\def\case#1/#2{{\textstyle{#1\over#2}}}
\def\9{\phantom 0}      
\renewcommand\linebreak{\unskip\break} 
\newcommand{\alt}{\mathrel{\raisebox{-.6ex}{$\stackrel{\textstyle<}{\sim}$}}}
\newcommand{\agt}{\mathrel{\raisebox{-.6ex}{$\stackrel{\textstyle>}{\sim}$}}}
\def\lsim{\alt}
\def\gsim{\agt}

\begin{document}
{\hspace*{\fill}{\vbox{\begin{flushright} \bf UCD-97-9 \\ April, 1997
\end{flushright}}}\\}
\bigskip
\title{A Simplified Summary of 
Supersymmetry~\thanks{To appear in {\it Future High Energy Colliders}, 
proceedings of the ITP Symposium, U.C. Santa Barbara, 
October 21-25, 1996, AIP Press.
Also presented at the Aspen Winter Conference on High Energy Physics, January
1997, Aspen, CO.}
}

\author{John F. Gunion$^*$}
\address{$^*$Davis Institute for High Energy Physics, Department of Physics,\\
University of California at Davis, Davis CA 95616}

\maketitle
\thispagestyle{empty}

\begin{abstract}
I give an overview of the motivations for
and theory/phenomenology of supersymmetry. 
\end{abstract}

\section*{Introduction}

The overview consists of three parts. Namely,
\begin{itemize}
\item WHY do theorists find supersymmetry so attractive?
\item WHAT do we expect to see experimentally?
\item HOW do we go about observing what is predicted?
\end{itemize}
In each part, I will give only the barest outline of the relevant issues
and discussions. In order to keep the presentation simple, I will often
be less than precise in the technicalities. Well-known results will
not be referenced in detail. Further discussion and references
can be found, for example, in \cite{wessbagger,ross,haberkane,dine}.

\section*{WHY supersymmetry is attractive}

First, there are some very general aesthetic considerations.
\begin{itemize}
\item \susy\ is the only non-trivial extension of the Lorentz group 
which lies at the heart of quantum field theory;  
the simplest such extension is referred to as $N=1$ supersymmetry
and requires the introduction of a single (two-component)
spinorial (anti-commuting) dimension to space, the extra dimension(s)
being denoted $\theta$.
Taylor expansions of a superfield $\widehat\Phi(x,\theta)$
then take the form
$\widehat\Phi(x,\theta)=\phi(x)+\sqrt2(\theta\psi(x))+(\theta\theta)F(x)$
(higher orders in $\theta$ being zero),
implying an automatic association of a spin-1/2 field $\psi(x)$
with every spin-0 field $\phi(x)$ and vice-versa ($F(x)$ is 
an auxiliary field, \ie\ it does not represent a dynamically independent
field degree of freedom).
\item If \susy\ is formulated as a {\it local} symmetry,
then a spin-2 (graviton) field must be introduced, thereby leading 
automatically to (SUGRA) models in which gravity is unified with
the other interactions. Further, SUGRA 
reduces to general relativity in the appropriate limit.
\item \susy\ appears in superstrings.
\item Historically, adding the Lorentz group to quantum mechanics
required introducing an antiparticle for every particle;
why should history not repeat --- \susy\ + quantum field theory 
requires a `sparticle' for every particle.
\end{itemize}

Of course, since we have not detected any of the superpartners
of the known Standard Model (SM) particles, it is clear that
supersymmetry is a broken symmetry. Thus, it
is possible that supersymmetry could be irrelevant at the energy scales
where experiments can currently be performed. However,
there are many reasons to suppose that the superpartners have
masses below a TeV and, therefore, could be discovered anytime now.
\begin{itemize}
\item String theory solutions with a non-supersymmetric {\it ground} 
state are quite problematic, implying that all the physical states
of the theory should have similar masses aside from the
effects of supersymmetry breaking (which should be a perturbation
on the basic string theory solution).
\item 
\susy\ solves the hierarchy problem, \eg\ $m_{\rm Higgs}^2<(1\tev)^2$ 
(as required to avoid a non-perturbative $WW$ sector),
via spin-1/2 loop cancellation of spin-0 loop quadratic divergences:
\begin{equation}
m_H^2\sim [m_H^0]^2+\lam^2 (m_{\rm boson}^2-m_{\rm fermion}^2)\ln
{\Lambda^2\over \langle m_b^2,m_f^2\rangle},~~~(m_f\sim
\msusy)\,;
\label{quaddiv}
\end{equation}
$m_H^2$ can be small if $[m_H^0]^2$ is, {\it provided} $m_f^2\lsim (1\tev)^2$.
\item
\susy\ implies gauge coupling unification (at $\mgut\sim {\rm few}\times
10^{16}\gev$) if $\msusy\lsim 1-10\tev$ and there is nothing
(other than complete SU(5) representations) between $\msusy$ and $\mgut$.
Of course there are some qualifications to this statement.
\begin{trivlist}
\item
[a)] Perturbative unification requires that the number of families (only
complete families lead to unification) must be $\leq 4$ \cite{4fam}.
\item
[b)] There is a possible difficulty in the string theory context
in that $\mgut<\mstring$ if we accept the perturbative
result that $\mstring=\mplanck/\sqrt{8\pi}\sim 2\times 10^{18}\gev$. 
However, it has been emphasized~\cite{lykken} that in non-perturbative
approaches it may be possible for $\mstring$ to be substantially lower,
in principle even as low as a TeV.
\item
[c)] Unification takes place only if there are
exactly 2 Higgs doublets (+ singlets)
below $\mgut$. This is also the minimal Higgs field
content required to give both up and down quarks masses and
to guarantee anomaly cancellation.
\item
[d)] The precise scale $\msusy$ preferred for exact unification depends
upon the precise value of $\alpha_s(\mz)$.  For currently accepted
values of $\alpha_s(\mz)\lsim 0.118$, an effective $\msusy$ value above 
$1\tev$ is seemingly preferred although other subtle issues could
alter this preference \cite{alphasbagger}.
\end{trivlist}
\item Electroweak symmetry breaking (EWSB) occurs automatically for
the simplest universal boundary conditions at $\mgut$ by virtue
of the fact that the mass-squared parameter for the Higgs
field coupled to the top-quark ($m_{H_2}^2$) is driven
negative (during evolution from $\mgut$
down to $\mz$)
by the large top-quark Yukawa contribution to its renormalization
group equation (RGE). The associated symmetry
breaking occurs very naturally at an energy scale
in the vicinity of $\mz$ if $m_{H_2}^2<(1-2\tev)^2$ at $\mgut$.
\item If the supersymmetric partner of the QCD gluon has mass below
$\sim 1\tev$, then most GUT boundary conditions imply
that the lightest supersymmetric particle (LSP) will
have mass on the $100\gev$ scale and would interact weakly
and have other properties that make it a natural candidate
for the cold dark matter of the universe. However, this is only
true if this LSP is essentially stable. In certain variants
of supersymmetry, to be discussed later, this is not the case.
\end{itemize}

\section*{WHAT we expect to see}
\begin{trivlist}
\item [A)] Sparticles: in the minimal supersymmetric model (MSSM)
every normal SM particle has its supersymmetric counterpart.
\begin{eqnarray}
\bigl[u,d,c,s,t,b\bigr]_{L,R}~~
\bigl[e,\mu,\tau\bigr]_{L,R}~~\bigl[\nu_{e,\mu,\tau}\bigr]_L
& ~~~~g~~~~ & \underbrace{\wpm,\hpm}
~~~\underbrace{\gam,Z,H_1^0,H_2^0}\nonumber \\
\bigl[{\tilde u},{\tilde d},{\tilde c},{\tilde s},{\tilde t},{\tilde
b}\bigr]_{L,R}~~\bigl[\tilde e,\tilde\mu,\tilde\tau\bigr]_{L,R}~~
\bigl[\tilde \nu_{e,\mu,\tau}\bigr]_L
 & ~~~~{\tilde g}~~~~& ~~~~\chitil_{1,2}^{\pm}~~~~~~~~~~\chitil_{1,2,3,4}^0
\nonumber
\end{eqnarray}
The quark, lepton and neutrino partners are the spin-0 squarks, sleptons and
sneutrinos; the partner of the gluon is the spin-1/2 gluino; and
the partners of the charged (neutral) vector bosons and Higgs bosons
are the spin-1/2 charginos (neutralinos). Often the latter can
be approximately separated into the spin-1/2 bino and wino
gaugino partners of the U(1) $B$ and  SU(2) $W$ gauge fields and 
the higgsino partners of the Higgs fields. In other cases, these
states are strongly intermixed.

There is a possibly exact discrete symmetry of the theory, called R-parity,
such that SM particles have $R=+$ while the sparticle partners have $R=-$.
If R-parity is an exact symmetry then any physical process must always
involve an even number of sparticles, and the LSP, normally
the $\cnone$, will be stable against decay to SM particles.  
\item [B)] A {\bf very} special Higgs sector \cite{hhg}.
\begin{itemize}
\item For the minimal two-doublet
Higgs sector, the physical eigenstates comprise two CP-even scalars 
($\hl$, $\hh$), a CP-odd scalar ($\ha$), and a charged Higgs pair ($\hpm$).
\item \susy\ implies that the Higgs self couplings have strength of order $g$,
the SU(2) SM coupling, which has the consequence that $\mhl\leq 130\gev$
($\leq 150\gev$, if singlet Higgs fields are added).
\item For boundary conditions such that EWSB is an automatic consequence
of the RGE's, 
$\mhh\sim\mha\sim\mhpm> 200\gev$ is very probable, in which
case the $\hl$ will have properties very much like those
predicted for the SM Higgs ($\hsm$) in the minimal one-doublet SM, while
the $\hh,\ha$ decouple from $WW,ZZ$.
\end{itemize}
\end{trivlist}

Thus, there is little question as to what we should see, but the
very uncertain nature of supersymmetry breaking implies that there
is a great deal of uncertainty as to the exact mass scale
at which we should see the new sparticles and as to the new experimental 
signatures that will appear when sparticles are produced. 

There is one important general point. 
If R-parity is exact, the sparticles must be produced in pairs. (Limits
on R-parity violation suggest that single sparticle production is
at best very weak in any case.) Thus, in order to observe supersymmetric
particles at hadron machines
we must have significant $gg$ and/or $q\anti q$ luminosity at
$\rts_{q\anti q,gg}$ above (very roughly) $2\msusy$;
at an $\epem$ or $\mupmum$ collider the $\rts$ must exceed the
sum of the masses of the the two sparticles one hopes to observe.
Large masses ($\gsim 1\tev$) for some sparticles are certainly possible,
in which case a lepton collider with $\rts\sim 3-4\tev$ will be needed.

With regard to the Higgs sector, the $\hl$ is guaranteed to be light
and should be easily detected, perhaps even at LEP2 or the Tevatron.
If a light $\hl$ is not found, then we must abandon the
possibility of low-energy supersymmetry as we now understand it.
In the $\mha\sim\mhh\sim\mhpm> 200\gev$ decoupling limit,
the heavier Higgs must be pair produced, \eg\ $\epem,\mupmum\to
\hh\ha,\hp\hm$. Energy reach could again prove to be crucial.

\subsection*{Supersymmetry breaking}

The couplings of the complete complex of sparticles and particles are
simultaneously fixed by the superpotential, denoted $W$, which involves
products of superfields (with their particle and sparticle component fields). 
As a result, almost all couplings involving sparticles are 
related by `Clebsch-Gordon' factors to the couplings
of their SM counterparts. The only exception is the possible presence
in $W$ of R-parity violating ($\rslash$) couplings. I shall temporarily
ignore such couplings, but will return later to this subject.

Most of the uncertainty in phenomenology is related to the many
possible scenarios for supersymmetry breaking, which lead to many different
predictions for the masses (especially relative masses)
of the sparticles and for the detailed experimental signatures 
that will be present when they are produced. The main constraint
on supersymmetry breaking is that it should be `soft' in the sense that
it should not destroy the very attractive \susy\ solution to the naturalness and
hierarchy problems. The possible supersymmetry breaking terms in
the Lagrangian can then be enumerated.

\begin{itemize}
\item
gaugino masses: $M_i\lam_i\lam_i$ where $i=3$, 2 ,1 for SU(3), SU(2), U(1)
and $\lam_i$ denotes the spin-1/2 partner of the corresponding gauge field.
\item
scalar masses: \eg\
\begin{equation}
m_{H_1}^2|H_1|^2+m_{H_2}^2|H_2|^2+m^2_{({\tilde t,\tilde
b})_L}({\stop_L}^\star\stop_L+{\sbot_L}^\star\sbot_L)+
m_{\stop_R}^2{\stop_R}^\star\stop_R+m_{\sbot_R}^2{\sbot_R}^\star\sbot_R
\nonumber
\end{equation}
\item `A' terms: \eg\ $A_t\lam_t(\stop_LH_2^0-\sbot_LH_2^+)\stop_R^\star$.
\item `B': $B\mu(H_1^0H_2^0-H_1^+H_2^-)$, where $\mu$ is the parameter
appearing in the superpotential term $W\ni\mu\what H_1\what H_2$.
\end{itemize}
Altogether, including CP-violating phases,
the above comprise 105 independent and unknown (although limited
in magnitude if we are to maintain the naturalness of the model
and avoid a charge and/or color breaking ground state) 
parameters beyond the SM. The obvious question is whether
the possibility of so many a priori unknown parameters is a good
or bad thing.
\begin{itemize}
\item Bad: If we want to know ahead of time exactly what to look for, then
it is bad in that the uncertainties associated with so many parameters 
imply a very large range of phenomenological possibilities.
\item Good: If we want to be confident that we will learn something from what
we observe, then the existence of so many a priori unknown parameters is good.
In particular, by evolving the low-energy parameters up to $\mgut$
one can hope to uncover $\mgut$-scale boundary conditions that imply
an underlying organization for the 105 parameters
that can be associated with an attractive and theoretically compelling 
GUT/string model.
\end{itemize}
The lesson for machine builders and experimentalists is that being
prepared for a wide range of
possibilities as to HOW? we see and fully explore \susy\ is a necessary evil.
The rest of the talk reviews some of the many possibilities
discussed to date.

\section*{How to look for supersymmetry}

It will be important to check that the Higg sector fits within
the supersymmetric model constraints and to learn whether
it contains more than the minimal two doublet fields,
in particular whether or not there are additional singlet Higgs fields.
Direct discovery of sparticles will be even more important.
We would hope to eventually observe all the sparticles of the theory.
I give a short description of Higgs phenomenology and then turn
to sparticle phenomenology.

\subsection*{Detecting the \susy\ Higgs bosons}

\begin{figure}[h]
\epsfxsize=3.5in
\centerline{\epsffile{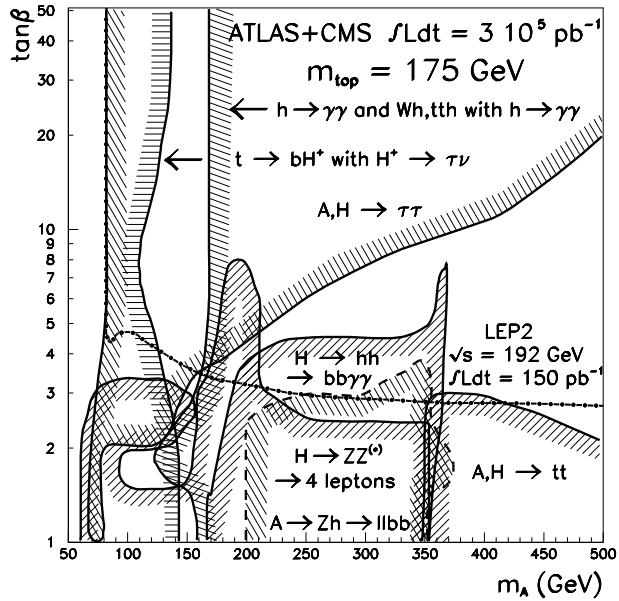}}
\vspace{10pt}
\caption[fake]{\baselineskip=0pt
Higgs discovery contours ($5\sigma$) in the parameter space of the 
minimal supersymmetric model for ATLAS+CMS at the LHC: $L=300\fbi$.
\Twoloop\ radiative corrections 
to the MSSM Higgs sector are included
assuming $\mstop=1\tev$ and no squark mixing. 
From Ref.~\protect\cite{latestplots}.}
\label{mssmhilum}
\end{figure}

It has been clearly established that for a \susy\ Higgs
sector consisting of only the minimal two-doublets
at least one of the \susy\ Higgs
bosons will be detectable at the large hadron collider (LHC) and
at the planned $\rts\sim 500\gev$ next linear lepton collider (NLC).
In the $\mha> 2\mz$ decoupling limit, it is always the $\hl$
whose observability is guaranteed.  Detection of the $\hh,\ha,\hpm$
is not guaranteed. In particular, at the NLC $\hh\ha$ and $\hp\hm$
pair production is not kinematically allowed if $\mha\gsim \rts/2$.
Also, at the LHC there is a region, see Fig.~\ref{mssmhilum},
in the standard $(\mha,\tanb)$
parameter space (which specifies the tree-level properties of 
the Higgs bosons) characterized by moderate $\tanb\gsim 3$ and $\mha> 200\gev$
such that only the $\hl$ will be detectable.

The two most urgent and best-motivated questions are:
\begin{itemize}
\item Is discovery of one Higgs boson guaranteed if singlet Higgs fields
are added to the minimal two doublet fields? This question
is particularly important in light of the fact that string theories
typically lead to a Higgs sector with extra singlets.
\item What is required in order to discover the $\hh,\ha,\hpm$
of the MSSM?
Here, the region of concern is the $\mha> 200\gev$ parameter
region that is natural when EWSB is automatically broken by
virtue of the RGE's.
\end{itemize}
Both questions have been explored in the literature and I summarize
the conclusions to date. 
\begin{itemize}
\item
It can be demonstrated \cite{kimoh,kot,KW,ETS} that 
at least one of the light CP-even Higgs bosons ($\h_{1,2,3}$)
of a two-doublet plus
one singlet Higgs sector will be observed at the NLC. This result
follows from the fact that if the lightest ($\h_1$) has weak $ZZ$
coupling then (one of) the heavier ones must actually be almost as light
and have substantial $ZZ$ coupling (and thus be discoverable
in the $\epem\to\zstar\to Z\h$ production mode). This result
probably extends to the inclusion of several singlet Higgs fields.
\item
At a $\mupmum$ collider, direct $s$-channel $\mupmum\to \h_1$ 
production is guaranteed to be visible \cite{bbghunpub}.
Only the (predicted) $(\mupmum\h_1)$ coupling is needed; the $\h_1$ can
be decoupled from $ZZ,WW$ without affecting its detectability.
A scan search in the $\leq 150\gev$ region is required with 
$\Delta E_{\rm beam}\lsim0.01\%$.
\item
At the LHC, discovery of at least one Higgs boson is no
longer guaranteed if a singlet Higgs field is present in
addition to the minimal two doublet fields \cite{ghm}. The
no-discovery ``holes'' in parameter space are never terribly large,
but are certainly not insignificant when $\tanb$ is in the moderate
$\tanb\sim 3-10$ range.
\item
At an $\epem$ collider, the only means for detecting 
any of the $\hh,\ha,\hpm$ when $\mha\gsim\rts/2$ 
(so that $\hh\ha,\hp\hm$ pair
production is forbidden) is to run the collider in
the $\gam\gam$ collider mode \cite{ghgamgam,bbc}.
In this way, single $\gam\gam\to\hh,\ha$
production can be observed for $\mha\lsim 0.8\rts$ provided high 
integrated luminosity (\eg\ $L\sim 200\fbi$) is accumulated.
There is still no certainty that a $\gam\gam$ collider facility will
be included in the final NLC plans, nor regarding the luminosity
that would be available.
\item
At a $\mupmum$ collider, direct $\mupmum\to\hh,\ha$ production 
will be observable for any $\mha$ between $\rts/2$ (the pair
production limit) and $\rts$ provided that
$\tanb\gsim 3$ (below this the LHC will find the $\hh,\ha$, see
Fig.~\ref{mssmhilum})
and that $L=100\fbi$ (for $\rts\sim 500\gev$) is available
for the $\rts/2\to\rts$ scan \cite{bbgh}.
It must be possible to maintain high
instantaneous luminosity throughout the scan region (perhaps requiring
two cheap storage rings designed for optimal luminosity in different $\rts$
ranges).
\item
At both the $\epem$ and $\mupmum$ collider, it is envisioned that
the machine will be upgraded to increasingly large $\rts$. Once
$\rts\gsim 2\mha$, 
$\hh\ha$ and $\hp\hm$ pair production will be observable.  It has
been shown that this will be true even if these Higgs bosons have
substantial decays to sparticles \cite{gk,fengmoroi}. 
The hierarchy motivations for
supersymmetry suggest that $\mha$ will certainly lie below $1-1.5\tev$.
A strong form of naturalness suggests $\mha\lsim 500\gev$ \cite{anderson}.
\end{itemize}
If a SM-like $\hl$ is observed, the most precise determination of all
its properties requires {\it both} $\rts=500\gev$ running at the NLC
{\it and} an $s$-channel scan of the resonance peak in $\mupmum\to\hl$
collisions at $\rts\sim\mhl$ \cite{snowmasssummary}. 
This argues strongly for having
{\it both} types of machine, especially since very substantial $L=200\fbi$
accumulated luminosity is needed at {\it both} machines 
in order to achieve good precision for a model-independent
determination of all the Higgs couplings and its total width.
Once the $\hh$, $\ha$ and/or $\hpm$ have been detected (whether
in pair production or single production at the muon collider), 
the relative branching
ratios for $\hh,\ha,\hpm$ decays to different types of channels can
be measured and very strong constraints will be placed on a host
of the important parameters of the \susy\ model, and thus
on the GUT boundary conditions \cite{gk,fengmoroi}. 
At a muon collider,
$\mupmum\to \hh,\ha$ studies would require substantial $L$ 
at a $\rts$ that is almost certain to be significantly different than that
employed for $\mupmum\to\hl$ studies. Perhaps more than one muon collider
will turn out to be needed.

\subsection*{Detecting the sparticles}

The phenomenology of sparticles is determined by
the nature and source
of supersymmetry breaking and whether or not R-parity is broken. A selection
of possibilities is discussed below. A useful recent review covering
some of the following theoretical material is \cite{amundsonsusy96}.

\subsubsection{{Minimal Supergravity (mSUGRA)}}

This is the simplest and most fully investigated model.  It assumes
that the boundary conditions are set at $\mgut$, that there is
a desert in between $\mgut$ and $\msusy$, and that R-parity is exact.
Further, at $\mgut$ the supersymmetry breaking parameters listed earlier
are taken to be universal and to have no CP-violating relative phases.
Universal boundary conditions are natural~\footnote{This
assumes a sufficiently simple form for the Kahler potential.}
in the picture where \susy-breaking ($\susyslash$)
arises in a hidden sector and is only communicated
to the visible sector at the GUT/string scale 
via interactions involving gravity (which of
course knows nothing about quantum numbers).
The result is a model specified by just five parameters and one sign.
The GUT boundary conditions are:
$$m_{H_1}=m_{H_2}=m_{\sq_i}=m_{\slep_i}\ldots=m_0,~~~
M_3=M_2=M_1=\mhalf,$$
$$A_{ijk}=A_0,~~~ |\mu|,~~~\sign(\mu),~~~ B_0.$$
Beginning with these boundary conditions at $\mgut$, evolution to low energies
$\lsim \msusy$ yields some simple and important results.
\begin{trivlist}
\item[$-$] $M_3:M_2:M_1\sim \alpha_3:\alpha_2:\alpha_3\sim 7:2:1$,
leading to a similar ratio of the ($\overline{\rm MS}$)
gluino to wino to bino masses.~\footnote{The gluino
pole mass is substantially larger than its $\overline{\rm MS}$ mass.} 
\item[$-$] Approximate squark degeneracy is maintained
for the first two generations, implying acceptably small FCNC.
\item
[$-$] The third family stops are normally significantly mixed and split, 
and the lightest squark will be the lighter $\stopone$ eigenstate.
\item[$-$] EWSB is automatic, as described earlier --- 
it is convenient to trade the parameters $|\mu|$ and $B$ 
for $\tanb$ and the (known) value of $\mz$.
\item[$-$] $|\mu|$ and $\mha\sim\mhh\sim\mhpm$ are typically large
($> 200\gev$) unless $\tanb\gg 1$;
\item[$-$] In the usual case where $\mu> M_2$, one finds the following:
the LSP is the (stable) $\cnone\simeq {\wtil B}$ with $\mcnone\sim M_1$ 
and is a good (cold) dark matter candidate; the lightest chargino ($\cpmone$)
and 2nd lightest neutralino ($\cntwo$) 
are approximately SU(2) charged and neutral winos
and have similar mass $\sim M_2$; 
the heavier chargino ($\cpmtwo$) and neutralinos
($\chitil^0_{3,4}$) are charged and neutral higgsinos with masses $\sim |\mu|$.
\item[$-$] If $m_0^2\gg\mhalf^2$, then squarks, sleptons, \etc\ are heavier
than the lighter (gaugino) $\chitil$'s and are approximately degenerate;
\item[$-$] If $m_0^2\ll\mhalf^2$, then $\mslep<\msq$  and
$\mstopone$ is distinctly smaller than the other $m_{\sq_i}$;
charged sleptons can be lighter than most gauginos and it is possible
(not included in later discussions) that the LSP could be the $\snu_\tau$.

\end{trivlist}

The experimental signatures are determined by how the sparticles decay.
In mSUGRA, the above-outlined mass hierarchy
$m_{\chitil^0_{3,4}}\sim m_{\chitil^\pm_2}>
m_{\chitil^0_2,\chitil^{\pm}_1}>\mcnone$ implies that most
signatures will result from chain decays ---
\eg\ 
$$
\gl\to q\anti q\cpmone\to
q\anti q+
\left\{ 
\begin{array}{l} 
\ell^\pm\nu\cnone \\ q\anti q\cnone 
\end{array}
\right.
\,.
$$
(It is important to note that $\gl$ decay leads with equal
probability to either $\ell^+$ or $\ell^-$.)
Since masses and mass differences are both substantial,
events will be characterized by fairly energetic jets and leptons and by
large $\etmiss$ from the very weakly-interacting, stable final $\cnone$'s.

\noindent $\diamondsuit$ \underline{Tevatron and LHC}

Extensive Monte Carlo studies have determined the region
of mSUGRA parameter space for which direct discovery of sparticles
will be possible. Cascade decays lead to events with
jets, missing energy, and various numbers of leptons.
The Tev33 option at Fermilab will
cover \cite{tev33msugra} 
the most natural \cite{anderson} portion of parameter space.
The maximum reach at the LHC is in the $1\ell+{\rm jets}+\etmiss$
channel; one will be able to discover squarks and gluinos
with masses up to several TeV \cite{bcptmsugra}. Some particularly important
types of events are the following.
\begin{itemize}
\item
$pp\to \gl\gl\to \mbox{jets}+\etmiss$ and $\ell^{\pm}\ell^{\pm}$+jets+$\etmiss$,
the latter being the like-sign dilepton signal \cite{likesign}.
The mass difference $\mgl-\mcpmone$ can be determined 
from jet spectra end points \cite{likesign,susyprec,bartlsusy96}, 
while $\mcpmone-\mcnone$ can be determined from
$\ell$ spectra end points in the like-sign 
channel \cite{likesign,susyprec,bartlsusy96} ---
an absolute scale for $\mgl$ can be estimated ($\pm 15\%$) by separating
the like-sign events into two hemispheres corresponding to the two $\gl$'s
\cite{likesign}, by a similar separation in the jets+$\etmiss$
channel \cite{bcptmsugra}, or variations thereof \cite{susyprec,bartlsusy96}.
\item $pp\to \cpmone\cntwo\to(\ell^{\pm}\cnone)(Z^*\cnone)$, which yields
a trilepton + $\etmiss$ final state when $Z^*\to \ell^+\ell^-$;
$\mcntwo-\mcnone$ is easily determined if
enough events are available \cite{bcpttri}.
\item $pp\to\slep\slep\to 2\ell+\etmiss$, detectable
at the LHC for $\mslep\lsim 300\gev$ \cite{bcptmsugra}, \ie\ the region
of parameter space favored by mixed dark matter cosmology \cite{bbcos}.
\item Squarks will be pair produced and, for $m_0\gg\mhalf$,  would lead to
$\gl\gl$ events with two extra jets emerging from the 
primary $\sq\to q\gl$ decays.
\end{itemize}

\noindent $\diamondsuit$ \underline{NLC} 

Important discovery modes include the following \cite{nlcreport}.
\begin{itemize}
\item $\epem\to \cpone\cmone\to (q\anti q\cnone~\mbox{or}~\ell\nu \cnone)+
(q\anti q\cnone~\mbox{or}~\ell\nu \cnone)$;
$\mcpmone$ and $\mcnone$ will be well-measured
using spectra end points and beam energy constraints.
\item $\epem\to \slep\slep\to (\ell\cnone)(\ell\cnone),(\nu\cmone)(\anti
\nu\cpone),\ldots$; masses will be well-measured.
\end{itemize}
The $\sq$'s and (if $m_0$ is big) $\slep$'s can be 
too heavy for pair production at $\rts=500\gev$.
If $m_0\sim 1-1.5\tev$ (the upper limit allowed by naturalness), 
then $\rts\gsim 2-3\tev$
is required. It could be that such energies will be more easily
achieved in $\mupmum$ collisions than in $\epem$ collisions.

For certain choices of the mSUGRA parameters, 
it can happen that the phenomenology is much more `peculiar' than
the canonical situation described above. One such choice of boundary
conditions is that termed the  
{\bf `Standard' Snowmass 96 Point} \cite{bartlsusy96}:
$$m_0=200\gev,~~~ \mhalf=100\gev,~~~ A_0=0,~~~ \tanb=2,~~~\mu<0\,.$$
These boundary conditions predict that the gluino mass is approximately the
same as the average squark mass; only the $\stopone$ and $\sbot_L$
are lighter than the $\gl$. Consequently, $\gl\to \sbot_L b$ 
almost 100\% of the time. Also, masses are not very large: \eg\ $\mgl=285\gev$,
$m_{\sbot_L}=266\gev$. At the LHC, 
there will be millions of spectacular events with $\gl\gl\to
bb\sbot_L\sbot_L\to\ldots$. At the NLC, all
$\sq\sq$ thresholds are $>\rts=500\gev$, but
$\slep\slep$ and $\cpone\cmone,\chitil_{1,2}^0\chitil_{1,2}^0$ 
pair processes are all kinematically allowed.

There are particular choices for mSUGRA boundary conditions
that have strong theoretical motivation in the context
of strings and/or supergravity. These include:
\begin{trivlist}
\item[$\clubsuit$] No-Scale \cite{noscale}: $\mhalf\neq0$, $m_0=A_0=0$;
\item[$\clubsuit$] Dilaton or Dilaton-Like \cite{ibanez}:  
$\mhalf=-A_0=\sqrt 3 m_0$ 
\end{trivlist}
Here the dilaton is the string `modulus' field associated
with coupling constant strength in string models. To many,
the dilaton-like boundary conditions appear to be
particularly worthy of being taken
seriously. To deviate significantly from
dilaton-like boundary conditions in Calabi-Yau and Orbifold models
requires going to an extreme in which the dilaton has nothing, or almost
nothing, to do with supersymmetry breaking.
The dilaton and no-scale models have many common features. Most importantly,
the small (or zero) $m_0$ compared to $\mhalf$ implies that
sleptons are light, which in turn leads to
excellent LEP2, TeV33, LHC, and NLC leptonic signals for supersymmetry
\cite{nsdil}, the only uncertainty being the overall rates
as determined by the overall mass scale, $\mhalf$. ($\tanb$ and
$\sign(\mu)$ are the only other free parameters in these models.)

A quite different set of mSUGRA boundary conditions is that motivated by
the assumption that there is hidden-sector dynamical \susy\ breaking 
without gauge singlets. In this case it is natural for all
dimension-3 \susy-breaking operators in the low-energy theory
to be very small, \ie\ $\mhalf,A_0\sim 0$ (and possibly $B_0\sim0$ as well)
\cite{bkn}.
Such boundary conditions, in combination with existing experimental
constraints, imply that the gluino would be lighter than the
lightest neutralino \cite{fm}, and would tend to emerge as part
of a relatively long-lived gluon-gluino bound state (denoted $R^0$) 
with mass $\sim 1.4\gev$ (according to lattice calculations).
Remarkably, this scenario cannot yet be absolutely excluded by
accelerator experiments \cite{lightgl}. 
In addition, it is possible for the photino in such a model to have
the necessary properties to be a dark matter candidate \cite{lightglcos}. 
Ongoing experiments and analyses will be able
to exclude this scenario in the near future. For example,
for such boundary conditions the lightest chargino must have $\mcpmone<\mw$
and should be discovered at LEP2 once substantial
luminosity is accumulated at $\rts\sim 190\gev$, despite the non-canonical
nature of predicted signals.  Searches for hadrons containing gluinos
could also provide strong constraints \cite{lightglhad}.

\subsubsection{Beyond mSUGRA: Non-universality}

Despite the very attractive nature of the mSUGRA boundary conditions,
it is certainly possible to find motivation for
many possible sources of non-universality. The two general classes
of non-universality are gaugino mass non-universality
and scalar mass non-universality. Ref.~\cite{nonuniversal}
provides a useful review and references. Due to lack of space I discuss
only (and very briefly) gaugino mass non-universality.
Models characterized by such non-universality include:~\footnote{Readers
not familiar with string terminology can simply take
these specifications as names.}
\begin{itemize}
\item The O-II orbifold string model in which all matter fields 
lie in the $n=-1$ untwisted sector and \susy-breaking is
dominated by the overall size modulus (not the dilaton).
\item $F$-term supersymmetry breaking with $F\neq$ SU(5) singlet, leading to
$\call \sim  {\langle F \rangle_{ab}\over \mplanck}\lam_a\lam_b\,,$
where $\lam_{a,b}$ ($a,b=1,2,3$) are the gaugino fields.
If $F$ is an SU(5) singlet then
$\langle F \rangle_{ab}\propto c\delta_{ab}$ and
we get standard universality, $M_1=M_2=M_3$ (at $\mgut$).
But, more generally $F$ can belong to a non-singlet SU(5) representation:
$F\in ({\bf 24}{\bf \times} 
{\bf 24})_{\rm symmetric}={\bf 1}\oplus {\bf 24} \oplus {\bf 75}
 \oplus {\bf 200}\,.$
which implies that 
$\langle F \rangle_{ab}=c_a\delta_{ab}$, with $c_a$ depending
on the representation (an arbitrary superposition of representations
is also possible).
\end{itemize}                                      
The $M_{3,2,1}$ for these different cases are compared in Table~\ref{masses}.

\begin{table}[h]
\caption{\baselineskip=0pt $M_a$ at $\mgut$ and $\mz$
for the four $F_{\Phi}$ irr. reps. 
and in the $\delgs\sim -4$ O-II model.
From Ref.~\protect\cite{nonuniversal}.}
\label{masses}
\begin{center}
\begin{tabular}{|c|ccc|ccc|}
\hline
\ & \multicolumn{3}{c|} {$\mgut$} & \multicolumn{3}{c|}{$\mz$} \cr
$F$ 
& $M_3$ & $M_2$ & $M_1$ 
& $M_3$ & $M_2$ & $M_1$ \cr
\hline 
${\bf 1}$   & $1$ &$\;\; 1$  &$\;\;1$   & $\sim \;7$ & $\sim \;\;2$ & 
$\sim \;\;1$ \cr
${\bf 24}$  & $2$ &$-3$      & $-1$  & $\sim 14$ & $\sim -6$ & 
$\sim -1$ \cr
${\bf 75}$  & $1$ & $\;\;3$  &$-5$      & $\sim \;7$ & $\sim \;\;6$ & 
$\sim -5$ \cr
${\bf 200}$ & $1$ & $\;\; 2$ & $\;10$   & $\sim \;7$ & $\sim \;\;4$ & 
$\sim \;10$ \cr
\hline
 $\stackrel{\textstyle O-II}{\delgs=-4}$ & $1$ & $\;\;5$ & ${53\over 5}$ & 
$\sim 6$ & $\sim 10$ & $\sim {53\over5}$ \cr
\hline
\end{tabular}
\end{center}
\end{table}

I have room for only a few remarks regarding how the
phenomenology changes as a function of boundary condition scenario.
I focus on the most extreme changes. See Ref.~\cite{nonuniversal} for
more details and references.
\begin{itemize}
\item The general relations that
$\mcnone\sim {\rm min}(M_1,M_2)$ and $\mcpmone\sim M_2$ imply
that $\mcpmone\simeq\mcnone$ (both are winos)
in the ${\bf 200}$  and O-II scenarios, where $M_2\lsim M_1$.
Some important consequences of this result are \cite{cdg}:
\begin{itemize}
\item
At the NLC, $\epem\to \cpone\cmone$ would be very
hard to see since the (invisible) $\cnone$ would take all of the $\cpmone$
energy in the $\cpmone\to\ell^{\pm}\nu\cnone,q^\prime\anti q\cnone$ decays.  
One must employ $\epem\to \gam\cpone\cmone$.
\item
At the LHC, the like-sign dilepton signal (coming again from
$\cpmone\to\ell^{\pm}\nu\cnone$ decays) would be very weak.
In the O-II model, $\mgl\sim \mcpmone\simeq\mcnone$
means that the jets from $\gl\to q\anti q\cnone$ decay would also be soft;
the standard jets+$\etmiss$ signature would be much
weaker than normal and the maximum $\mgl$ for which \susy\
could be discovered would be smaller than in the usual case.
\item Dark matter phenomenology would be substantially altered.
In particular, the very close degeneracy $\mcpmone\simeq\mcnone$ implies
very similar $\cpmone,\cnone$ densities (due to very similar Boltzmann factors)
at freeze-out which in turn leads to $\cnone\cpmone$
`co-annihilation' that greatly reduces relic dark matter ($\cnone$) density. 
\end{itemize}
\item Generally speaking, if $\mgl$ is not large then distinguishing
between the five scenarios would be quite easy.  For example,
if we keep $m_0$ and $\mgl$ the same as at the Snowmass 96 point
described earlier, there would be many millions of $\gl\gl$
pair events at the LHC, and the different scenarios would
lead to the following very different event characteristics \cite{nonuniversal}:
\begin{itemize}
\item $\bf 1$: $\etmiss$, $\ell^+\ell^-$, 4$b$'s;
\item $\bf 24$: $\etmiss$, no $\ell^+\ell^-$, 8$b$'s, with
2 pairs having mass $\mhl$.
\item $\bf 75$: traditional cascade chain decay signals;
\item $\bf 200$: $\etmiss$, 4$b$'s, no leptons;
\item O-II: $\gl\to \cpmone,\cnone$ + soft and 
$\cpmone\to \cnone$ + very soft; soft jet cuts required to observe.
\end{itemize}
\end{itemize}

\subsubsection{R-parity violating models}

I next consider the possibility that R-parity is violated.  R-parity violation
can come in two different forms:

\begin{itemize}
\item Hard: that is explicit terms in the superpotential of the form
$W_{\rslash}=\lam_{ijk}(\what L_i\what L_j)\what{\anti E_k} +
\lam^\prime_{ijk}(\what L_i\what Q_j)\what{\overline D_k} 
+\lam^{\prime\prime}_{ijk}\what{\anti U_i} 
\what{\anti D_j}\what{\overline D_k}$.
\item Soft: $W_{\rslash}=\mu_i \what L_i \what H_1$
\end{itemize}
I make a few brief remarks regarding the former.
(Phenomenology for the latter is reviewed in Ref.~\cite{rviolhemp}.)
First, $\lam,\lam^\prime\neq0$ implies lepton-number violation,
while $\lam^{\prime\prime}\neq0$ implies baryon-number violation.
Both cannot be present; if they were, the proton lifetime would be short.
Otherwise, current constraints are not terribly strong \cite{rviollimits}: 
$\lam$'s $\lsim 0.1-0.01$ for superparticle masses in the $>100\gev$ range. 
Phenomenologically, the crucial point is that
unless the $\lam$'s are {\it very, very} small, the LSP
$\cnone$ will decay inside the detector. Thus, the signals
for supersymmetry will no longer involve missing
energy associated with the $\cnone$. Impacts on phenomenology
are substantial.

\bigskip
\noindent At the LHC \cite{likesignrviol}: 
\begin{itemize}
\item
If $\lampp\neq 0$, then $\cnone\to 3j$. The large jet backgrounds
imply that we would need to rely on the like-sign dilepton signal.
For universal boundary conditions, 
this signal turns out to be sufficient for supersymmetry discovery
out to $\mgl$ values somewhat above 1 TeV.
However, if the leptons are very soft, as occurs if
$\mcpmone\sim\mcnone$ (as in the $\bf 200$ and O-II models) then
the discovery reach would be much reduced --- 
the combination of $\rslash$
and $M_2\lsim M_1$ would be a bad scenario for the LHC.
\item If $\lam\neq 0$, $\cnone\to \mu^\pm e^\mp\nu,e^\pm e^\mp \nu$,
and there would be many
very distinctive multi-lepton signals.
\item If $\lampr\neq 0$, $\cnone\to \ell 2j$ and again there would
be distinctive multi-lepton signals.
\end{itemize}
At the NLC:
\begin{itemize}
\item
Even $\epem\to\cnone\cnone\to \underbrace{(3j)(3j)}_{\lampp\neq 0}, 
~~\underbrace{(2\ell\nu)(2\ell\nu)}_{\lam\neq 0}, 
~~\underbrace{(\ell 2j)(\ell 2j)}_{\lampr\neq0}$ could
yield an observable \susy\ signal.
\end{itemize}
At HERA:
\begin{itemize}
\item
Squark production via R-parity violating
couplings  could be an explanation for the HERA anomaly
at high $x$ and large $Q^2$ \cite{many}.
For example, if $\lam^\prime_{113}\sim 0.04-0.1$,
a lepto-quark--like signal from
$e^+ d_R\to \wtil t_L\to e^+ d$ would be detected if $m_{\wtil t_L}\lsim
220\gev$. (We focus on the top squark since it can easily be the
lightest squark in supergravity models.)
\end{itemize}
Finally, if R-parity is violated then supersymmetry will no longer
provide a source for dark matter, the LSP no longer being stable.
One would have to turn to neutrinos (which, however, only provide
a source of hot dark matter, whereas some cold dark matter
also seems to be needed).

\subsubsection{Event-Motivated Models}

A number of specific supersymmetry breaking parameter choices have been
proposed in order to explain particular anomalous events seen 
at the Tevatron and at LEP. Here, I only have room to list some
of these and make a few remarks.  In all
cases the $\mgut$-scale boundary conditions would be required to be
non-universal (assuming the usual desert between low energy and $\mgut$).

\bigskip
\centerline{
\underline{CDF $ee\gam\gam$ event = $\sel\sel$, $\cpone\cmone$, $\ldots$
production?}}
\bigskip

Let us focus on the $\sel\sel$ explanation. One proposed explanation
of this event \cite{eegamgamkane} as $\sel\sel$ production requires that
the one-loop decay
$\cntwo\to \gam \cnone$ dominate over all tree level decays
{\it and} that $\br(\sel\to e\cntwo)\gg\br(\sel\to e\cnone)$.
If both are true, then the process
$$\wtil e\wtil e \to (e\cntwo) (e\cntwo)\to (e\gam\cnone)(e\gam\cnone)$$
would explain the observed event provided masses are also appropriate.
For the above decays to be dominant requires
$\cntwo=\wtil \gam~~\mbox{and}~~ \cnone=\mbox{higgsino}$, which in turn
is only the case if $M_2\sim M_1~~\tanb\sim 1~~~|\mu|<M_{1,2}.$
This, in combination with the observed kinematics of the event
and a cross section consistent with its having been detected,
implies masses
for all the neutralinos and charginos that are mostly in the $50-150\gev$ range.
For such masses there should be a large number of other equally
distinctive events in the $L\sim 100\pbi$ of Tevatron data currently
being analyzed and clear signals should also emerge at LEP when
run at $\rts\sim 190\gev$.
At the Tevatron, $L\sim 100\pbi$ implies
$N(2\ell+X+\etmiss)\geq30$, $N(\gam\gam+X+\etmiss)\geq 2$,
$N(\ell\gam+X+\etmiss)\geq 15$, $N(2\ell\gam+X+\etmiss)\geq 4$,
$N(\ell2\gam+X+\etmiss)\geq 2$, $N(3\ell+X+\etmiss)\geq 2$.
At LEP190 with $L=500\pbi$, one finds
$N(2\ell+X+\etmiss)\geq 50$, $N(2\gam+X+\etmiss)\geq 3$.
In the above, $X=$additional leptons, photons, jets.
Finally, I note that the small $\mu$
value needed implies that we must have soft mass non-universality
in the form $m_{H_1}^2\neq m_{\sq,\slep}^2$.

\bigskip
\centerline{ \underline{CDF and D0 `di-lepton top events' 
that don't look like top events}}
\bigskip

There are two events in the CDF di-lepton top sample and one event
in the D0 sample that have very low probability, in terms of their
kinematics, to be top events.
It has been proposed \cite{barnetthall} that these could be
from $\sq\to q\chitil\to q (\nu\slep)$ with $\slep\to \ell\cnone$,
where $\chitil=\wtil W_3,\wtil W^\pm$. For this explanation
to work in terms of kinematics and cross section requires:
$\mgl\simeq 330$, $\msq\simeq 310$,
$\mslepl\simeq 220$, $m_{\snu_\ell}\sim 220$,
$\mslepr\simeq 130$, $\mu\sim-400$,
$M_1\simeq 50$, and $M_2\simeq 260$.
Note that these values are inconsistent with the previous
$ee\gam\gam$ event explanation. An abundance of other signals should
be hidden in the full Tevatron data set for such parameter choices.

\bigskip
\centerline{
\underline{The Four-Jet `Signal' at ALEPH}}
\bigskip

The 4-jet signal at ALEPH is well-known. It has $M+M^\prime\sim 106\gev$ and 
$M-M^\prime\sim10\gev$ (\ie\ $M\sim 58,M^\prime\sim 48\gev$),
significant `rapidity-weighted' jet charge and no $b$ jets.
One interpretation \cite{4jetrviol} is as
$\sell\selr$ production with $\sell,\selr\to 2j$ via $\rslash$ coupling
$W_{\rslash}\ni \lam^\prime_{ijk}(\what L^i\what Q^j)_L\what{\overline D^k_R} $.
$M_1\lsim 100\gev$ is required for the needed $\sigma\sim 1-2\pb$ cross
section level, and is also helpful in suppressing
$\sell\sell,\selr\selr$ production.
To suppress (unobserved) $\sel\snue$ production we need the largest $\msnue$
possible, which occurs for $\tanb\sim 1$. 
For $\sell\to\anti u_jd_k$ 
to dominate over the more standard $\sell\to ee\selr$ 
decays via $\cnone$ exchange
requires $\lam_{1jk}\gsim \mbox{few}\times 10^{-4}$, which
is entirely consistent with known bounds.
In this picture, the $\selr$ must decay by mixing,
$\selr\to\sell\to\anti u_jd_k$ rather than via $\selr\to ee\anti ud$ 
from virtual $\cnone,\sell$ exchange; the required
mixing angle, $\sin\phi\gsim 10^{-4}$, is easily accommodated.
One finds that a displaced vertex might be observable.
The absence of $\wtil \mu$ and $\wtil\tau$ pair signals requires
that these states have
substantially heavier masses than the $\sel$, as would only be possible if
slepton masses are non-universal. Finally, lots of signals would
be expected at LEP2: 
for $M_1=100\gev$, $\msnue=58\gev$ and $\rts=186\gev$, 
$$\sigma(\sell\selr,\sell\sell,\selr\selr,\snuel\snuel,\cnone\cnone)=
1.33,0.23,0.2,0.29,0.62\pb$$

\subsubsection{Models with gauge-mediated supersymmetry breaking}

Underlying all our previous discussions has been the implicit
assumption that \susy\ is almost certainly an exact local symmetry
(as in strings) which, if not accidental, should be spontaneously
broken in a `hidden sector'. The \susy-breaking is then fed to the
ordinary superfields in the form of an effective Lagrangian, $\call_{\rm eff}$.
The mass expansion parameter for $\call_{\rm eff}$ is 
the mass scale of the sector responsible for the
communication between the hidden sector and the ordinary superfields.
For gravity-mediated communication (as appropriate in SUGRA/superstring
theories), this mass scale is most naturally $\sim\mplanck$ and 
one arrives at the
$\call_{\rm eff}\sim  {\langle F \rangle\over \mplanck}\lam_a\lam_a$
expression given earlier. For gaugino masses of order $\mw$,
this requires $\langle F \rangle\equiv \msusyslash^2\sim\mw\mplanck\sim
(10^{11}\gev)^2$. 

The gravity-mediated scenario is certainly
very attractive but has its doubters, primarily based on
the fact that keeping FCNC phenomena at a sufficiently small level
is not guaranteed.
Indeed, although gravity being `flavor blind' suggests universal
soft scalar masses, terms that violate universality
are certainly possible in the Kahler potential
and (since not forbidden by symmetry)
are generated at some level by radiative corrections, even 
if not present at tree-level. Further, a certain amount of FCNC
and lepton-number violation  can arise during evolution
from $\mstring\to\mgut$, although, as noted earlier,
$\mstring\sim\mgut$ is also possible non-perturbatively.
A final point of concern is that
EWSB via the RGE's is not guaranteed for all possible boundary condition
choices. All these problems can be resolved if the scale
of supersymmetry breaking, $\msusyslash$, is much lower.

The specific and popular models that incorporate this idea are
the gauge-mediated supersymmetry breaking (GMSB) models.
In these models, $\mplanck$ above is replaced by $M$,
the mass scale of a new messenger sector,
and $\msusyslash^2\sim \mw M$ can be much lower --- 
$\msusyslash\sim 10-100\tev$
is often discussed.  The GMSB models have rather few parameter,
at least in current incarnations.  They lead to dramatic signals, but
the signals could well be somewhat different than at first anticipated
because of issues related to the mass scale of supersymmetry breaking.
I briefly outline the basics \cite{basicgmsb,dine}.

In the standard GMSB models there are three sectors.
\begin{trivlist}
\item[I:] First, there is the \susy-breaking sector, sometimes
called the `secluded' sector,  containing hidden
particles that interact via strong gauge
interactions which cause supersymmetry breaking characterized by
a scale $\sqrt F$. The \susy-breaking is then fed via {\it two-loops} 
in the strong interaction into
\item[II:] the `messenger' sector, which contains a 
SU(3)$\times$SU(2)$\times$U(1) (but not
necessarily SU(5)) singlet superfield $\widehat S$
with non-zero vacuum expectation values for both its scalar component
and its $F$-term component ($\vev{S}\neq0$ and $\vev{F_S}\neq 0$).
The two-loop communication
between the two sectors implies that $\vev{F_S}\sim (\alpha_m^2/16\pi^2) F$,
where $\alpha_m$ characterizes the strength of the gauge interactions
that are responsible for the two-loop communication between the \susy-breaking
sector and the messenger sector. In addition to $\widehat S$,
the messenger sector must contain some messengers that transform
under SU(3), SU(2) and U(1). In order to maintain actual
gauge-coupling unification, these messengers should form a complete
anomaly-free GUT representation (\eg\ $\bf 5+{\bf \overline 5}$
with messengers $\widehat q,\widehat{\anti q}$ and $\widehat\ell,\widehat{\anti \ell}$
superfield triplets and doublets in the SU(5) case). 
These messenger superfields must communicate with $\widehat S$; 
a typical superpotential
might be $W=\lam_1\what S\what q \what{\anti
q}+\lam_2\what S\what\ell\what{\anti\ell}$. 
The mass scales
of the component boson ($b$) and fermion ($f$) fields 
of these messenger superfields
are given very roughly by 
\begin{equation}
M\equiv m_{f_{\rm mess.}}\sim \lam \vev{S};~~~ m^2_{b_{\rm mess.}}\sim \left(
\begin{array}{cc} \lam^2\vev{S}^2 & \lam\vev{F_S} \\ \lam\vev{F_S} &
\lam^2\vev{S}^2\\ \end{array} \right)
\label{mfbforms}
\end{equation}
(where $\lam$ is the typical $\lam_{1,2}$),
implying that $\lam\vev{S}^2>\vev{F_S}$ is 
required to avoid a negative mass-squared eigenvalue.
Ratios of $m^2_b$ eigenvalues $<30$ (no fine tuning) suggests
${\lam\vev{S}^2/\vev{F_S}}\gsim 1.05$.
\item[III:]
These messengers communicate \susy-breaking
to the normal superfield sector according to a scale characterized
by $\Lambda\equiv \vev{F_S}/\vev{S}$ 
(with $M/\Lambda=\lam\vev{S}^2/\vev{F_S}>1$ required,
as noted above). After integrating
out the heavy messengers, and assuming that $\widehat S$ is an SU(5) singlet,
the masses of the particles important to low-energy phenomenology are
as follows.
\begin{itemize} 
\item
The gauginos acquire masses at one-loop given by
\begin{equation}
M_i(M)=k_iN_{5,10}\, g\left({\Lambda\over M}\right) 
{\alpha_i(M)\over 4\pi}\Lambda\,,
\label{inomasses}
\end{equation}
where $k_2=k_3=1$, $k_1=5/3$, and 
$N_{5,10}$ is the number of ${\bf 5}+{\bf \anti 5}$ plus
three times the number of ${\bf 10}+{\bf \anti 10}$
messenger representations. $N_{5,10}\leq 4$ is required to avoid Landau poles.
The gaugino mass ratios are the same as found
for universal boundary conditions at $\mgut$ in mSUGRA models.
\item The squarks/sleptons acquire masses-squared at two-loops: $m_i^2(M)=$
\begin{equation}
{\textstyle
2\Lambda^2N_{5,10} f\left({\Lambda\over M}\right)
\left[c_3\left({\alpha_3(M)\over
4\pi}\right)^2+c_2\left({\alpha_2(M)\over 4\pi}\right)^2+{5\over 3}\left({Y\over
2}\right)^2\left({\alpha_1(M)\over 4\pi}\right)^2\right]\,,}
\label{scalarmasses}
\end{equation}
with $c_3=4/3$ (triplets) $c_2=3/4$ (weak doublets), $Y/2=Q-T_3$. 
Degeneracy among families is 
broken only by effects of order quark or
lepton Yukawa couplings, implying no FCNC problems in the first two
generations.
\end{itemize}
In the above, $g(\Lambda/M),f(\Lambda/M)\to 1$ for  $M/\Lambda$ not
too near 1. For $\Lambda=M$, $g(1)=1.4, f(1)=0.7$. 
\begin{itemize}
\item Spontaneous \susy-breaking leads to a goldstone fermion,
the goldstino, $G$. (In local SUSY, $G$ is the longitudinal component of 
the gravitino.) $G$ acquires mass determined by $F$, 
the goldstino decay constant, where $F$ is the largest $F$-term 
vev (here, the $F$ of the secluded, true supersymmetry breaking sector):
\begin{equation}
m_G={F\over \sqrt 3\mplanck}\sim 2.5
\left({\sqrt{F}\over 100\tev}\right)^2~{\rm eV}\,;
\label{mgform}
\end{equation}
As sketched earlier, $\sqrt F\sim 10^8 \tev$ is appropriate in
the usual mSUGRA/string-motivated models; the $G$ is then fairly massive
and, since it is also very weakly interacting, it is
irrelevant to low-energy physics.  In GMSB models, the much
smaller $\sqrt F\sim 100-1000\tev$ values envisioned imply
that the $G$ will be the lightest supersymmetric particle.
In this case, a very crucial constraint is that 
$m_G$ be $\lsim 1\kev$ to avoid overclosing the universe; \ie\
$\sqrt F\lsim 2000\tev$ is required for consistency with cosmology.
\end{itemize}
\end{trivlist}

Some useful observations affecting the phenomenology
of a GMSB model are the following.
\begin{itemize}
\item FCNC \etc\ problems are 
solved since the gauge interactions are flavor-blind
and, thus, so are the soft terms. 
\item  As already noted, if $\what S$ is an SU(5) singlet, then we obtain
the usual mSUGRA prediction: $M_3:M_2:M_1=7:2:1$ at scale $\sim\mz$.
\item Eq.~(\ref{inomasses}) implies 
\begin{equation}
\Lambda={\vev{F_S}\over \vev S}\sim{80\tev\over N_{5,10}}
\left({M_1\over 100\gev}\right)\,.
\label{m1norm}
\end{equation}
\item Eqs.~(\ref{inomasses}) and (\ref{scalarmasses})
imply (for $g=f=1$)
\begin{equation}
\msq:\mslepl:\mslepr:M_1=11.6:2.5:1.1:\sqrt{N_{5,10}}\,,
\label{ratios}
\end{equation}
which in turns implies that the lightest of the standard sparticles 
(referred to as the next-to-lightest supersymmetric particle,
denoted NLSP  --- the goldstino being the LSP)
is the $\wtil B$ for $N_{5,10}=1$ or the $\slepr$ for $N_{5,10}\geq2$.
\item $\mu$ and $B$ do not arise from the GMSB ansatz and are `model-dependent',
but EWSB driven by negative $m_{H_2^0}^2$ turns out to be completely automatic.
\item 
The couplings of the $G$ are fixed by a supersymmetric Goldberger-Treiman
relation:
$$\call=-{1\over F}j^{\mu\alpha}\partial_\mu G_\alpha+h.c.$$
where $j$ is the supercurrent connecting a SM particle to its superpartner.
For $\sqrt F$ values in the range of interest, this coupling is very weak,
implying that all the superparticles other than the NLSP undergo
chain decay down to the NLSP. The NLSP finally decays to the $G$:
\eg\ $\wtil B\to \gam G$ (and $ZG$ if $m_{\wtil B}>\mz$) 
or $\slepr\to \ell G$. The $c\tau$ for
NLSP decay depends on $\sqrt F$; \eg\ for $N_{5,10}=1$
\begin{equation}
(c\tau)_{\cnone=\wtil B\to \gam G}\sim 130
\left({100\gev\over M_{\wtil B}}\right)^5
\left({\sqrt F\over 100\tev}\right)^4\mu {\rm m}
\label{ctauform}
\end{equation}
If $\sqrt F\sim 2000\tev$ (the upper limit from cosmology),
then $c\tau\sim 21$m for $M_{\wtil B}=100\gev$; 
$\sqrt F\sim 100\tev$ implies a short but vertexable decay length. 

\end{itemize}

If the final decay of the NLSP occurs rapidly ($\sqrt F\sim 100\tev$), 
then there are many highly
observable signatures for a GMSB model \cite{commongmsb}. For example,
GMSB with the $\cnone$ being the NLSP could explain the Tevatron $ee\gam\gam$
event as  $\sel\sel\to e\cnone e\cnone\to ee \gam \gam GG$ 
(the $G$'s yielding $\etmiss$). However, I will now argue that $\sqrt F$
is most naturally very near the $\sqrt F\sim 2000\tev$
upper bound, in which case rather few of the NLSP's decay
inside the detector. In fact, in their current incarnation
the GMSB models have a significant problem of scale related
to the value of $f\equiv F/\vev{F_S}$. As noted earlier,
the communication between
the true supersymmetry breaking sector and the messenger sector
occurs at two-loops, implying $f\sim \left({g_m^2\over 16\pi^2}\right)^{-2}\sim
2.5\times 10^4/g_m^4$
where $g_m$ refers to the gauge group responsible for \susy-breaking, whereas
Eqs.~(\ref{mfbforms}) and (\ref{m1norm}) imply
\begin{equation}
f ={F\over \vev{F_S}}=
\left({\sqrt F\over 2000\tev}\right)^2 {625 \lam N_{5,10}^2\over (M/\Lambda)}
 \left({100\gev\over M_1}\right)^2\,.
\label{fform}
\end{equation}
The problem is that a value as large as 
$f\sim 2.5\times 10^4/g_m^4$ is generally rather inconsistent with 
basic phenomenological constraints 
if (as hoped) $g_m\lsim 1$. 
\begin{itemize}
\item
To illustrate, consider
$\lam\sim 1$, $g_m=1$ (\ie\ $f=2.5\times 10^4$) and
$M/\Lambda\sim 1$. (Recall $M/\Lambda> 1$ is required;
the choice $M/\Lambda\sim 1$ minimizes the scale problem.)
If $\sqrt F\sim 2000\tev$, \ie\ as large as possible consistent with
$m_G\leq 1$ keV, then Eq.~(\ref{fform}) implies:
$M_1\sim 16\gev$ ($63\gev$) for $N_{5,10}=1$ ($N_{5,10}=4$). The former
is experimentally ruled out. The latter might
be acceptable, but for $N_{5,10}=4$ the NLSP is the $\slepr$ with
$\mslepr\sim 35\gev$, see Eq.~(\ref{ratios}), 
which is ruled out by $Z$ data. Although the inconsistency
in this latter case is not very bad and could
be resolved by modest increases in $\lam$ and/or $g_m$,
or corrections to our simple two-loop estimate for $f$, 
the $\slepr$ NLSP would appear as
a heavily ionizing track of substantial length in the detector
when $\sqrt F$ is as large as assumed.
Most probably such events would have been
observed at the Tevatron in the $\mslepr\lsim 100\gev$
range roughly consistent with $f\sim 2.5\cdot 10^4$.
\item A value of $\sqrt F\sim 100\tev$, as taken in many phenomenological
discussions and studies \cite{commongmsb}, 
is highly inconsistent with the two-loop
expectations for $f$. (This has also been
noted in Ref.~\cite{murayama}, where it was used
to motivate attempts to construct models in which the supersymmetry
breaking sector and the messenger sector are one and the same.)
For example, taking $\sqrt F=100\tev$,
$M_1=100\gev$, $M/\Lambda\sim 1$, $\lam\sim 1$ and 
$N_{5,10}=1$ ($N_{5,10}=4$) in Eq.~(\ref{fform}) results in $f=1.56$ ($f=25$).
\item If an acceptable model
with 1-loop communication between the \susy-breaking
sector and the messenger sector can be constructed,
we would predict  $f\sim 16 \pi^2/g_m^2\sim 160/g_m^2$.
For $g_m=1$ (\ie\ $f=16\pi^2$), $M_1=100\gev$, $M/\Lambda\sim 1$, 
$\lam\sim 1$ and $N_{5,10}=1$ 
($N_{5,10}=4$), Eq.~(\ref{fform}) 
yields $\sqrt F\sim 1000\tev$ ($\sqrt F\sim 250\tev$).
The associated NLSP lifetimes, see \eg\ Eq.~(\ref{ctauform}),
would lead to easily detected vertices in events where sparticles are produced.
Analysis of existing Tevatron data and forthcoming LEP2 data should
readily uncover supersymmetric signals.
\end{itemize}

In short, for models in which the \susy-breaking
sector communicates at two (or even one) loops
with the messenger sector, it seems to be 
inconsistent to take $\sqrt F\lsim 100\tev$. This implies that
the simple phenomenology in which the NLSP decays ($\cnone \to \gam G$ or
$\slepr\to\ell G$) almost immediately is not relevant, and
the GMSB explanation of the Tevatron $ee\gam\gam$
event would be very unlikely.
The large $\sqrt F$ values required for even a modicum of consistency
imply that experimentalists should be looking for events with  
vertices a substantial distance from the interaction point,
quite possibly in association with
heavily ionizing tracks.  In the most
natural case, where $\sqrt F\sim 1000-2000\tev$ and $\mcnone$ and $\mslepr$
are below $100\gev$, $c\tau$ ---
see \eg\ Eq.~(\ref{ctauform}) --- is typically many tens to several hundreds
of meters and only a fraction $\sim 2R/[\vev{\gam} c\tau]$
of the events will have at least one vertex inside a detector of radius $R$. 
\begin{itemize}
\item If the NLSP is the $\slepr$, the bulk of \susy\ events will have 
several heavily ionizing tracks in the detector, a small fraction
of which will suddenly terminate with the emission of a $\ell$.
Since there is surely no background to such events, this possibility
should be excludable using existing Tevatron data (given the significant
production rates associated with the low sparticle masses
required for scale consistency).
\item If the NLSP is the $\wtil B$, the bulk of events 
will not have a vertex inside the detector and will appear as typical
$\etmiss$ supersymmetry signal events;
high rates would be required to uncover such events. However,
the small, but significant (given the low sparticle masses), 
number of very distinctive events in which a 
photon (or $Z$) suddenly emerges in the 
middle of the electromagnetic or hadronic 
calorimeter should have a small
background, in which case only a few events would be required for discovery.  
It would seem that this GMSB scenario might
also be ruled out or confirmed with proper analysis of Tevatron, if not LEP2,
data. 
\end{itemize}
If the above kinds of events, consistent with 
the most natural $\sqrt F\sim 1000-2000\tev$ values are found, it will
be useful to construct `far-out' additions to the CDF and D0
detectors designed to reveal more of the decay vertices.

The cosmological implications/consistency of GMSB constitute an important issue
\cite{gmsbcosdim,gmsbcosmur},
but one that is far too complex to elaborate on significantly here.
I only note that if $\sqrt F\sim 100\tev$, as disfavored
by the scale consistency discussed above, then $m_G\sim 2.5$~eV implies
that the goldstino cannot make a significant contribution to the
dark matter of the universe; however, the messenger sector might contain
an appropriate dark matter candidate \cite{gmsbcosdim}.
In a  model with
$\sqrt F\sim 1000\tev$, as preferred by the scale arguments, 
the $G$ gives rise to a cosmologically significant abundance of warm
dark matter; unfortunately, it would be invisible in halo detection 
experiments \cite{gmsbcosdim}.

\section*{Conclusions}

If supersymmetry is discovered, it will be a dream-come-true for
both theorists and experimentalists.
For theorists, it would a a triumph of aesthetic principles, naturalness, \etc\
For experimentalists, it would be a gold mine of experimental signals
and analyses. As has always been the case in the past, the next step
in theory will require experimental guidance and input.
The many phenomenological manifestations
and parameters of supersymmetry imply that many years of experimental
work will be required before it will be possible to determine
the precise nature of supersymmetry breaking and the
associated boundary conditions. Our ultimate dream
is that, armed with this information,
we will be able to construct the `final' theory.

\section*{Acknowledgements}

This work was supported in part by the Department of Energy, by
the Davis Institute for High Energy Physics and by the Institute
for Theoretical Physics.  I am
grateful to B. Dobrescu and H. Murayama for helpful discussions.


\begin{references}

\bibitem{wessbagger} J. Wess and J. Bagger, 
{\it Supersymmetry and Supergravity} (Princeton
University Press, Princeton, 1983), and references therein.

\bibitem{ross} G. Ross, {\it Grand Unified Theories} (Frontiers
in Physics, Benjamin Cummings, 1984)

\bibitem{haberkane} H.E. Haber and G.L. Kane, \PR 117 75 1985 .

\bibitem{dine}  M. Dine, hep-ph/9612389, and references therein.


\bibitem{4fam} See, for example, J.F. Gunion, D.W. McKay and H. Pois,
\PRD D53 1616 1996 .

\bibitem{lykken}J. Lykken, \PRD D54 3693 1996 .

\bibitem{alphasbagger} D. Pierce, J.A. Bagger, K. Matchev and R.-J. Zhang,
hep-ph/9606211 and references therein.

\bibitem{hhg} Pedagogical treatments and
detailed references can be found in:
J.F. Gunion, H.E. Haber, G. Kane and S. Dawson,
{\it The Higgs Hunters Guide} (Frontiers in Physics, Addison Wesley, 1990);
{\it Perspectives on Higgs Physics}, ed. G. Kane 
(World Scientific Publishing, 1993);
J.F. Gunion, A. Stange, and S. Willenbrock,
{\it Weakly-Coupled Higgs Bosons}, preprint UCD-95-28 (1995),
to be published in {\it Electroweak Physics and Beyond the Standard Model}
(World Scientific Publishing),
eds. T. Barklow, S. Dawson, H. Haber, and J. Siegrist.

\bibitem{latestplots}
D. Froidevaux, F. Gianotti, L. Poggioli,
E. Richter-Was, D. Cavalli, and S. Resconi, 
ATLAS Internal Note, PHYS-No-74 (1995).  

\bibitem{kimoh}
B.R. Kim, S.K. Oh and A. Stephan, 
{\it Proceedings of the 2nd International Workshop on
``Physics and Experiments with Linear $\epem$ Colliders''},
eds. F. Harris, S. Olsen, S. Pakvasa and X. Tata, Waikoloa, HI (1993), 
World Scientific Publishing, p.~860.

\bibitem{kot}
J. Kamoshita, Y. Okada and M. Tanaka, \PLB B328 67 1994 .

\bibitem{KW} S.F. King and P.L. White, preprint SHEP-95-27 (1995).

\bibitem{ETS} U. Ellwanger, M.R. de Traubenberg and C.A. Savoy,
\ZP C67 665 1995 .

\bibitem{bbghunpub} V. Barger, M. Berger, J.F. Gunion and T. Han,
unpublished.

\bibitem{ghm} J.F. Gunion, H.E. Haber and T. Moroi, hep-ph/9610337,
to be published in the proceedings of the 1996 DPF/DPB Summer Study
on {\it New Directions for High-Energy Physics} (hereafter
referred to as Snowmass 96).

\bibitem{ghgamgam}
J.F. Gunion and H.E. Haber, Proceedings of the 1990 DPF Summer Study on 
{\it High Energy Physics: Research Directions for the Decade}, 
editor E. Berger, Snowmass (1990), p. 206; and \PRD D48 5109 1993 .

\bibitem{bbc}
D. Borden, D. Bauer, and D. Caldwell, \PRD D48 4018 1993 .

\bibitem{bbgh} V. Barger, M. Berger, J. Gunion and T. Han, 
\PRL 75 1462 1995 ; hep-ph/9602415, to appear in Physics Reports.

\bibitem{gk}  J.F. Gunion and J. Kelly, hep-ph/9610495; 
a summary will appear in Snowmass 96.

\bibitem{fengmoroi} J. Feng and T. Moroi, hep/ph-9612333.

\bibitem{anderson} G. Anderson and D. Castano, \PLB B347 300 1995 ;
\PRD D52 1693 1995 ; \PRD D53 2403 1996 .

\bibitem{snowmasssummary} J.F. Gunion, L. Poggioli and R. Van Kooten,
to appear in Snowmass 96.

\bibitem{amundsonsusy96} J. Amundson \etal, hep-ph/9609374,
to appear in Snowmass 96.

\bibitem{tev33msugra} See, for example, H. Baer. C.-H. Chen, F. Paige
and X. Tata, \PRD D54 5866 1996 ; D. Amidei \etal, FERMILAB-PUB-96-082;
and references therein.

\bibitem{bcptmsugra} 
H. Baer, C.-H. Chen, F. Paige and X. Tata, \PRD D52 2746 1995 ;
\PRD D53 6241 1996 .

\bibitem{likesign} R.M. Barnett, J.F. Gunion, and H.E. Haber, \PLB B315 349
1993 ;

\bibitem{susyprec} I. Hinchliffe, F. Paige, M. Shapiro, 
J. Soderqvist and W. Yao, hep-ph/9610544.

\bibitem{bartlsusy96} A. Bartl \etal, to appear in Snowmass 96.

\bibitem{bcpttri} H. Baer, C.H. Chen, F. Paige and X. Tata, 
\PRD D50 4508 1994 .

\bibitem{bbcos} H. Baer and M. Brhlik, \PRD D53 597 1996 .


\bibitem{nlcreport} 

T. Tsukamoto, K. Fujii, H. Murayama, M. Yamaguchi and Y. Okada, \PRD D51 3153
1995 ;
H. Baer, R. Munroe and X. Tata, \PRD D54 6735 1996 .
For a review, see
{\it Physics and Technology of the Next Linear Collider:
a Report Submitted to Snowmass 1996}, SLAC Report 485.

\bibitem{noscale} See, for example, A. Lahanas and D. Nanopoulos, \PR 145 1
1987 .

\bibitem{ibanez} A. Brignole, L. Ibanez and C. Munoz, \NPB B422 125 1994 ;
hep-ph/9508258. See also V. Kaplunovsky and J. Louis, \PLB B306 269 1993 .

\bibitem{nsdil} 
H. Baer, J. Gunion, C. Kao and H. Pois, \PRD D51 2159 1995 ;
J. Lopez, D. Nanopoulos, X. Wang, and A. Zichichi, \PRD D52 142 1995 ;
J. Lopez, D. Nanopoulos and Z. Zichichi, \PRD D52 4178 1995 .


\bibitem{bkn} T. Banks, D. Kaplan and A. Nelson, \PRD D49 779 1994 .

\bibitem{fm} G. Farrar and A. Masiero, hep-ph/9410401.

\bibitem{lightgl} G.R. Farrar, \PRD D51 3904 1995 .

\bibitem{lightglcos} G.R. Farrar and E.W. Kolb, \PRD D53 2990 1996 .

\bibitem{lightglhad} G.R. Farrar, \PRL 76 4111 1996 .

\bibitem{nonuniversal} G. Anderson, C.H. Chen, J.F. Gunion, J. Lykken,
T. Moroi, Y. Yamada, hep-ph/9609457, to appear in Snowmass 96.

\bibitem{cdg} C.-H. Chen, M. Drees and J.F. Gunion, \PRL 76 2002 1996 ; 
\PRD D55 330 1997 .

\bibitem{rviolhemp} R. Hempfling, hep-ph/9609528, hep-ph/9702412.

\bibitem{rviollimits} See, for example,  V. Barger, G.F. Giudice,
and T. Han, \PRD D40 2987 1989 ;
S. Davidson, D. Bailey and B. Campbell, \ZP C61 613 1994 .


\bibitem{likesignrviol} An incomplete set of references is:
P. Binetruy and J.F. Gunion, proceedings of the {\it Workshop
on Novel Features of High Energy Hadronic Collisions}, Erice, Italy, 1988,
published in {\it Eloisatron: Heavy Flavors 1988}, p. 489;
H. Dreiner, M. Guchait and D.P. Roy, \PRD D49 3270 1994 ; 
H. Baer, C.-H. Chen and X. Tata, \PRD D55 1466 1997 ; A. Bartl \etal,
hep-ph/9612436.

\bibitem{many} An incomplete list is: D. Choudhury and S. Raychaudhuri,
hep-ph/9702392; G. Altarelli, J. Ellis, G.F. Giudice,
S. Lola and M. Mangano, hep-ph/9703276; H. Dreiner and P. Morawitz,
hep-ph/9703279; J. Kalinowksi, R. Ruckl, H. Spiesberger
and P.M. Zerwas, hep-ph/9703288; K.S. Babu, C. Kolda, J. March-Russell
and F. Wilczek, hep-ph/9703299; J. Hewett and T. Rizzo, hep-ph/9703337.

\bibitem{eegamgamkane} 
S. Ambrosanio, G.L. Kane, G.D. Kribs, S.P. Martin
and S. Mrenna, \PRD D55 1372 1997 .

\bibitem{barnetthall} R.M. Barnett and L.J. Hall, \PRL 77 3506 1996 ;
and hep-ph/9609313.

\bibitem{4jetrviol} M. Carena, G.F. Giudice, S. Lola and C. Wagner,
hep-ph/9612334.

\bibitem{basicgmsb} See, for example, 
M. Dine, A. Nelson and Y. Shirman, 
\PRD 51 1362 1995 ; M. Dine, A. Nelson, Y. Nir and Y. 
Shirman, \PRD D53 2658 1996 ; 
and references to earlier work therein.

\bibitem{commongmsb} 
See, for example, S. Dimopoulos, M. Dine, S. Raby
and S. Thomas, \PRL 76 3494 1996 ; 
S. Ambrosanio, G. Kane, G. Kribs, S. Martin and S. Mrenna, \PRL 76 3498 1996 ;
S. Dimopoulos, S. Thomas and J.D. Wells, \PRD D54 3283 1996 ; 
H. Baer, M. Brhlik, C.-H. Chen and X. Tata, \PRD D55 4463 1997 ; 
J. Bagger, K. Matchev, D. Pierce and R.-J. Zhang,
\PRL 78 1002 1997 .

\bibitem{murayama} N. Arkani-Hamed, J. March-Russell and H. Murayama,
hep-ph/9701286. See also, G. Dvali and M. Shifman, hep-ph/9612490.

\bibitem{gmsbcosdim} S. Dimopoulos, G. Giudice and A. Pomarol, \PLB B389 37
1996 .

\bibitem{gmsbcosmur}
A. de Gouvea, T. Moroi and H. Murayama, hep-ph/9701244.


\end{references}
\end{document}